\documentclass[12pt,preprint]{aastex}

\slugcomment{Submitted to The Astrophysical Journal Letters}
\lefthead{Agustsson \& Brainerd}
\righthead{Satellite Galaxies are Radially Aligned}

\begin{document}

\title{The Orientation of Satellite Galaxies: Evidence of Elongation
in the Direction of the Host}

\author{Ingolfur Agustsson \& Tereasa G. Brainerd}
\affil{Boston University, Institute for Astrophysical Research, 
725 Commonwealth Ave., Boston, MA 02215}
\email{ingolfur@bu.edu, brainerd@bu.edu}

\begin{abstract}
We use the fourth data release of the Sloan Digital Sky Survey to
investigate the orientations of 4289 satellite
galaxies with respect to their hosts.
The orientation of the satellites is inconsistent
with a random distribution at the 99.94\% confidence level, and
the satellites show a preference for elongation in the
direction of their hosts.  Further, on
scales $< 50$~kpc the major axes of the host galaxies and their satellites
are preferentially aligned.
Phrased in the terminology of weak lensing, the 
images of the satellites have a mean shear of 
$\gamma_T = -0.045 \pm 0.010$, averaged over scales $10~{\rm kpc} \le 
r_P  \le 50~{\rm kpc}$. 
In a galaxy--galaxy lensing study where
lenses and sources are separated solely on the basis of
apparent magnitude,  we estimate that on 
scales $\lesssim 250$~kpc  satellite galaxies account for 
between 10\% and 15\% of the objects that are identified as sources.
In such studies,
elongation
of the satellites will
cause a reduction
of the galaxy--galaxy lensing shear by of order 25\% to 40\%.
Hence, the elongation of satellite galaxies in the direction of their hosts is
a potentially important effect for precision studies
of galaxy--galaxy lensing, and argues
strongly in favor of the use of accurate photometric redshifts 
in order to identify
lenses and sources in future studies.
\end{abstract}


\section{Introduction}

Galaxy--galaxy lensing, hereafter GG lensing,
has become a premiere tool for constraining
the nature of the dark matter halos of galaxies (e.g., Brainerd
2004a, Brainerd \& Blandford 2002, and references therein). 
Recent investigations of GG lensing have  
moved beyond the most basic constraints on the nature of the ``average''
dark matter halo, demonstrating that there are real physical
differences between the halos of early--type and late--type galaxies,
and that the halos are non-spherical (e.g., Hoekstra et al.\ 2004;
Kleinheinrich et al.\ 2004; Sheldon et al.\ 2004; Mandelbaum et al.\
2005a; Seljak et al.\ 2005).  From its 
earliest days, however, GG lensing has been haunted by
the possibility that a number of
genuine satellite galaxies, in orbit about
the lenses,  could be mistakenly 
identified as
sources.  
If such satellites are randomly oriented with 
respect to the lenses, their presence introduces noise
in the measurement of the lensing signal.   If
the satellites are non--randomly oriented with respect to the lenses,
this alters the observed 
lensing signal compared to the true signal that would be measured in
the absence of such false sources.
Non--random orientations  
could be caused by tidal distortions at 
small distances from the host or, at larger distances,
by the tendency of galaxies to form in preferential alignment within
filaments (e.g., Catelan et al.\ 2000;
Croft \& Metzler 2000; Heavens et al.\ 2000;
Lee \& Pen 2000, 2001; Crittenden et al.\ 2001;
Brown et al.\ 2002; Jing 2002; Heymans et al.\ 2004; Mandelbaum et al.\
2005b)

Phillips (1985), Tyson (1985),
and Brainerd et al.\ (1996) used the clustering strength
of faint galaxies to place limits on contamination of the
GG lensing signal caused by
satellites and concluded that the contamination was sufficiently small
to be ignored.  Bernstein \& Norberg (2002), hereafter BN, found that 
on scales $< 500$~kpc, the mean 
tangential ellipticity of satellites in 
the Two Degree Field Galaxy Redshift Survey (2dFGRS; Colless et al.\ 2001,
2003)
was consistent with zero, 
and concluded that the contamination of the GG lensing
signal was $< 20$\%.  Hirata et al.\ (2004) used photometric redshifts
in an 
analysis of Sloan Digital Sky Survey data 
(SDSS\footnote{\url{http://www.sdss.org}};
Fukugita et al.\ 1996; Hogg et al.\ 2001;
Smith et al.\ 2002; Strauss et al.\ 2002; York et al.\ 2002)
and concluded that
on scales of 30$h^{-1}$~kpc to 446$h^{-1}$~kpc, the mean intrinsic
shear of satellite galaxies was consistent with zero and that the
contamination of the GG lensing signal due to satellites
was $\lesssim 15$\%.

Here we use the
fourth data release (DR4) of the SDSS (Adelman--McCarthy et al.\ 2005) to
revisit the question of whether satellite galaxies have a preferred
orientation with respect to their hosts.
Throughout we adopt $H_0 = 70$~km~sec$^{-1}$~kpc$^{-1}$,
$\Omega_{m0} = 0.3$, and $\Omega_\Lambda = 0.7$.

\section{Host and Satellite Galaxies in the SDSS DR4}

Hosts and satellites are selected by requiring: [1]
the hosts are relatively isolated and [2]
host--satellite pairs are nearby to one another
in terms of projected separation, $r_P$, and radial
velocity difference, $|dv|$.  Specifically, hosts must be 2.5 times
more luminous than any other galaxy that falls within $r_P \le 700$~kpc
and $|dv| \le 1000$~km~sec$^{-1}$.  Satellites must be
at least 4 times less luminous than their host, and must be located within
$r_P \le 250$~kpc and $|dv| \le 500$~km~sec$^{-1}$.  
We use
only galaxies with redshift confidence parameter
${\rm zconf} > 0.9$ and, to avoid systematics due to overlapping
isophotes, we use only those satellites that are located at radii
larger than three times the scale radius, $r_s$, of their host.
Lastly, we visually inspect the images of all candidate satellites
that are found within $r_p = 50$~kpc, and we reject those 
which have been misidentified as ``galaxies'' in the 
database.  Only 6.5\% of the candidate satellites within
50~kpc are rejected this way, 
and these consist of either a star or a bright blue knot that has been
misidentified as a small, faint galaxy.
Implementation of all our
criteria yields a final sample of
3180 hosts and 4289 satellites.  The median redshift of the hosts is
$z_{\rm med} = 0.058$.  Distributions of observed $r$--band apparent
magnitudes, the difference in observed $r$ magnitude, and
K--corrected colors are shown
in Fig.~1.  K--corrections were obtained from version 3.2 of
Michael Blanton's IDL 
code\footnote{\url{http://cosmo.nyu.edu/blanton/kcorrect/}} (e.g., Blanton et
al.\ 2003).
Since the satellites are large and
well--resolved, their images are not greatly affected by local anisotropies
in the PSF and, therefore, we make no corrections to their image shapes
below.

\begin{figure}
\centerline{
\scalebox{0.85}{%
\includegraphics{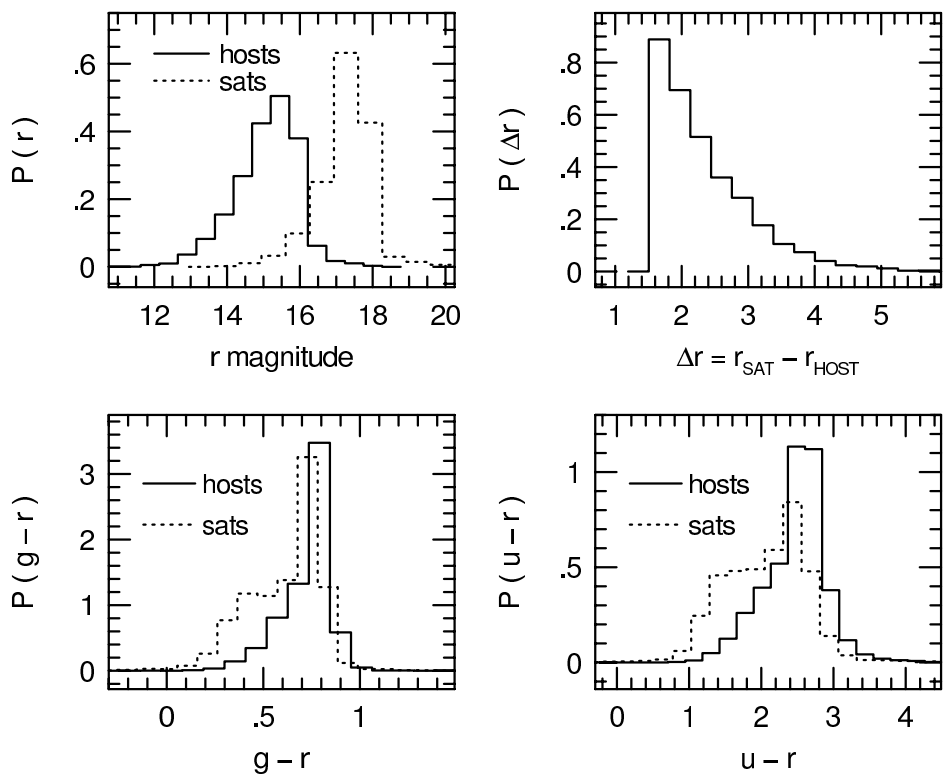}%
}
}
\caption{Top panels: 
Distribution of observed $r$ magnitude for the hosts and satellites
(left) and distribution of difference in observed
$r$ magnitude (right).  Bottom panels:
Distribution of K--corrected colors for the hosts and satellites.
}
\label{fig1}
\end{figure}

\section{Orientation of the Satellites}

We use the $r$-band position angles of the satellites to compute $\theta$,
the angle between the major axes of the satellites and
the direction vectors on the sky that connect the centroids of the
hosts and their satellites.  We restrict $\theta$ to
the range $\left[ 0^\circ , 90^\circ \right]$, where $\theta = 0^\circ$
corresponds to a radial orientation of the satellite in the direction of
its host and $\theta = 90^\circ$ corresponds to a tangential
orientation.
Shown in the top panels of Fig.~2 are the differential probability
distribution, $P(\theta)$, and continuous cumulative probability 
distribution, $P(\theta \le \theta_{\rm max})$, for the orientations
of the satellites.  The data in both panels are inconsistent
with random distributions.  
A $\chi^2$ test performed on $P(\theta)$
rejects the random distribution at the 99.93\% confidence level, while
a Kolmogorov--Smirnov (KS) test performed on $P(\theta \le \theta_{\rm max})$
rejects a random distribution at the
99.94\% confidence level.  From the top panels of Fig.~2, then, there
is a preference for the satellites to be elongated in the direction
of their hosts.

\begin{figure}
\centerline{
\scalebox{0.85}{%
\includegraphics{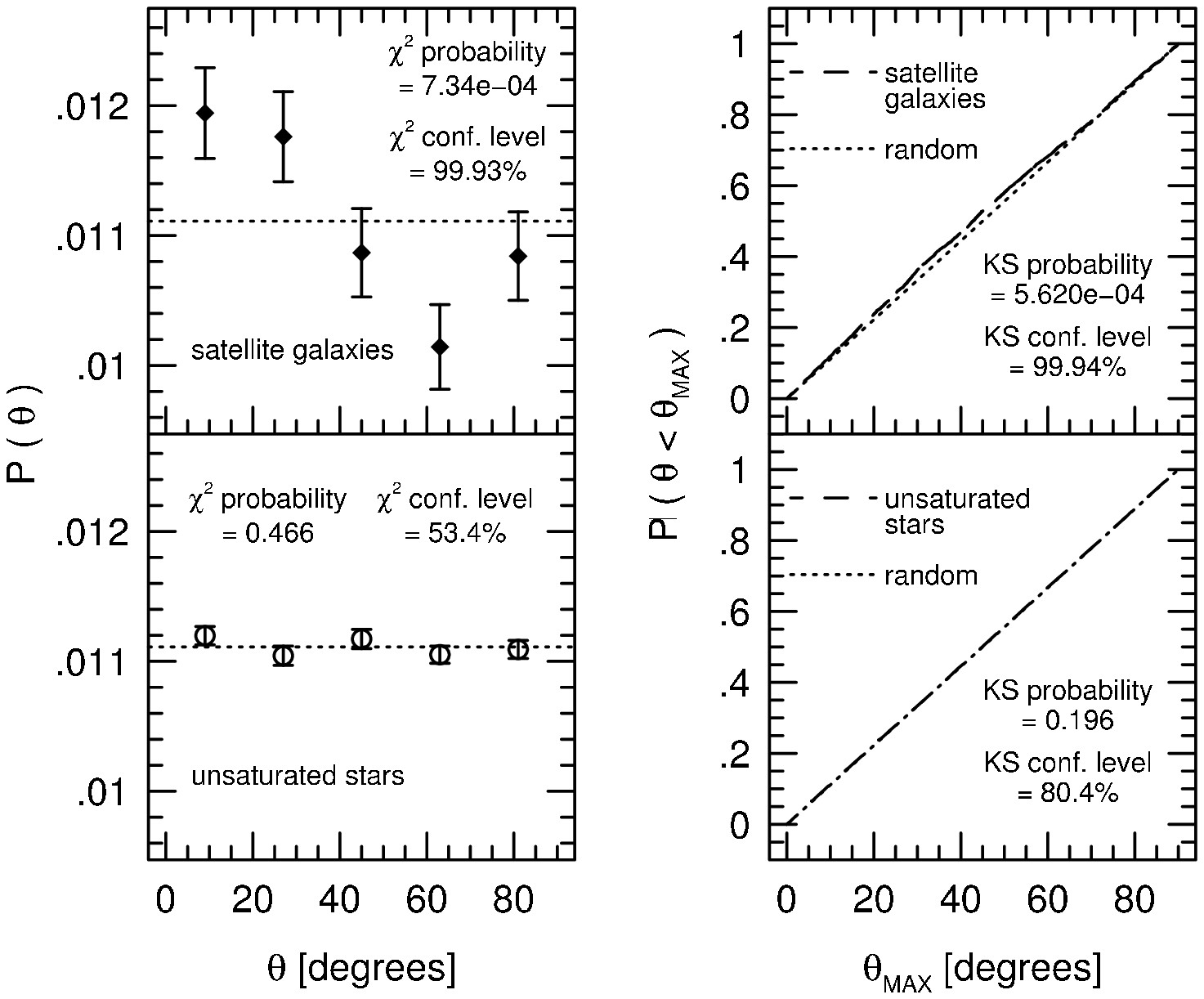}%
}
}
\vskip 0.1cm
\caption{
Left panels: Differential probability distribution
for the orientation of satellite galaxies (top) and 
stars with $16.5 \le r \le 18.5$ (bottom),
relative to the locations of the host galaxies.
Dotted lines show $P(\theta)$ for a random distribution.  
Formal rejection confidence levels from the $\chi^2$ test are
shown in the panels.
Right panels: Cumulative probability distribution for the
orientation of the satellite galaxies (top) and stars
(bottom).  Dotted lines show
$P(\theta \le \theta_{\rm max})$ for a random distribution.
Formal
rejection confidence levels from the KS test are shown in the panels.
}
\label{fig2}
\end{figure}

Shown in the bottom panels of Fig.~2 is a null test in which 
unsaturated stars with magnitudes similar to those of the 
satellite galaxies
are used
to compute $P(\theta)$ and
$P(\theta \le \theta_{\rm max})$.
The stellar sample consists of 92,489 stars in the SDSS
photometric database that have $16.5 \le r \le 18.5$ and are
found within projected radii  $3r_s \le r_p \le 250$~kpc of each host galaxy.
Here $r_s$ is again the scale radius of the host galaxy.  
Both $P(\theta)$ and $P(\theta \le \theta_{\rm max})$ for the stars 
are consistent
with random distributions and we are, therefore, confident that the
non--random orientation of the satellites shown in the
top panels of Fig.~2 is unlikely to be caused by systematics in the
imaging (e.g., drift--scanning, overlapping
image isophotes, and/or classical aberrations).
We specifically do not perform a null test that
is sometimes performed in GG lensing: the substitution of
the images of the hosts for those of the satellites.  That is,
if the ``lens'' galaxies are foreground
objects and the ``source'' galaxies are background
objects, the images of the sources should
be tangentially aligned with respect to the lenses but the
images of the lenses should be randomly oriented with respect to
the sources.  This 
pre--supposes that the centroids of the sources are distributed uniformly
around the lenses; however, this is not the case for our satellites.
Brainerd (2005) showed that satellites
are found preferentially close to the major axes of their hosts.  That is,
the major axes of the hosts point preferentially toward the
locations of their satellites (i.e., the hosts are radially
aligned along the direction vectors that connect the centroids of the
hosts with their satellites).  Indeed, the satellites in
our present sample have a mean location angle
of $ \left< \phi \right> = 42.4^\circ \pm 0.4^\circ$ 
relative to the major axes of their hosts, consistent with the 
results of Brainerd (2005). 

The combination of the results of Brainerd (2005) and our results in
Fig.~2 lead to the conclusion that host
galaxies and their satellites are intrinsically aligned on small
scales.  We quantify
this by computing a two--point correlation function of the shapes of the
hosts and their satellites:
$C_{\gamma \gamma} (r_P) \equiv \left< \vec{\gamma}_{h} \cdot
\vec{\gamma}_{s}^\ast \right>_{r_P}
$.  This is analogous to a function used in weak lensing to
measure 
correlated distortions in the images of lensed galaxies
as a function of the separation of the images on the sky (e.g., Blandford
et al.\ 1991). Here $\vec{\gamma}_h$ and $\vec{\gamma}_s$ are 
shape parameters for
the hosts and satellites, respectively, where $\vec{\gamma} \equiv \epsilon
e^{2i \varphi}$, $\varphi$ is the position angle of a galaxy, 
$a$ and $b$ are its major and minor axes, and $\epsilon \equiv
(a-b)/(a+b)$.
The mean, denoted by angle brackets, 
is computed over all pairs of hosts and satellites separated by
projected radii $r_P \pm 0.5 ~dr_P$.  
Host and satellite images that are uncorrelated
yield $C_{\gamma \gamma} (r_P) = 0$, while host and satellite images that
are aligned yield positive values of
$C_{\gamma \gamma} (r_P)$.  Solid circles in the top panel of Fig.~3 show 
$C_{\gamma \gamma} (r_P)$
for our hosts and satellites, where
it is clear that on scales $< 50$~kpc the images of the hosts and satellites
are aligned with each other.  On scales $> 50$~kpc, the images of the hosts
and satellites show no apparent correlation, consistent with the lack of
large scale
intrinsic alignment of SDSS galaxies reported by Mandelbaum et al.\
(2005b).  For comparison, open circles in the top panel of Fig.~2 show
$C_{\gamma\gamma}(r_P)$ computed using the shape parameters of the hosts
and the stars with $16.5 \le r \le 18.5$ that are nearby to the hosts.  
Unlike the images of the hosts and satellites,
the images of the hosts and nearby stars are uncorrelated, and we conclude
that the apparent correlation of host and satellite images on scales
$< 50$~kpc is unlikely to be caused by imaging systematics.

\begin{figure}
\centerline{
\scalebox{0.85}{%
\includegraphics{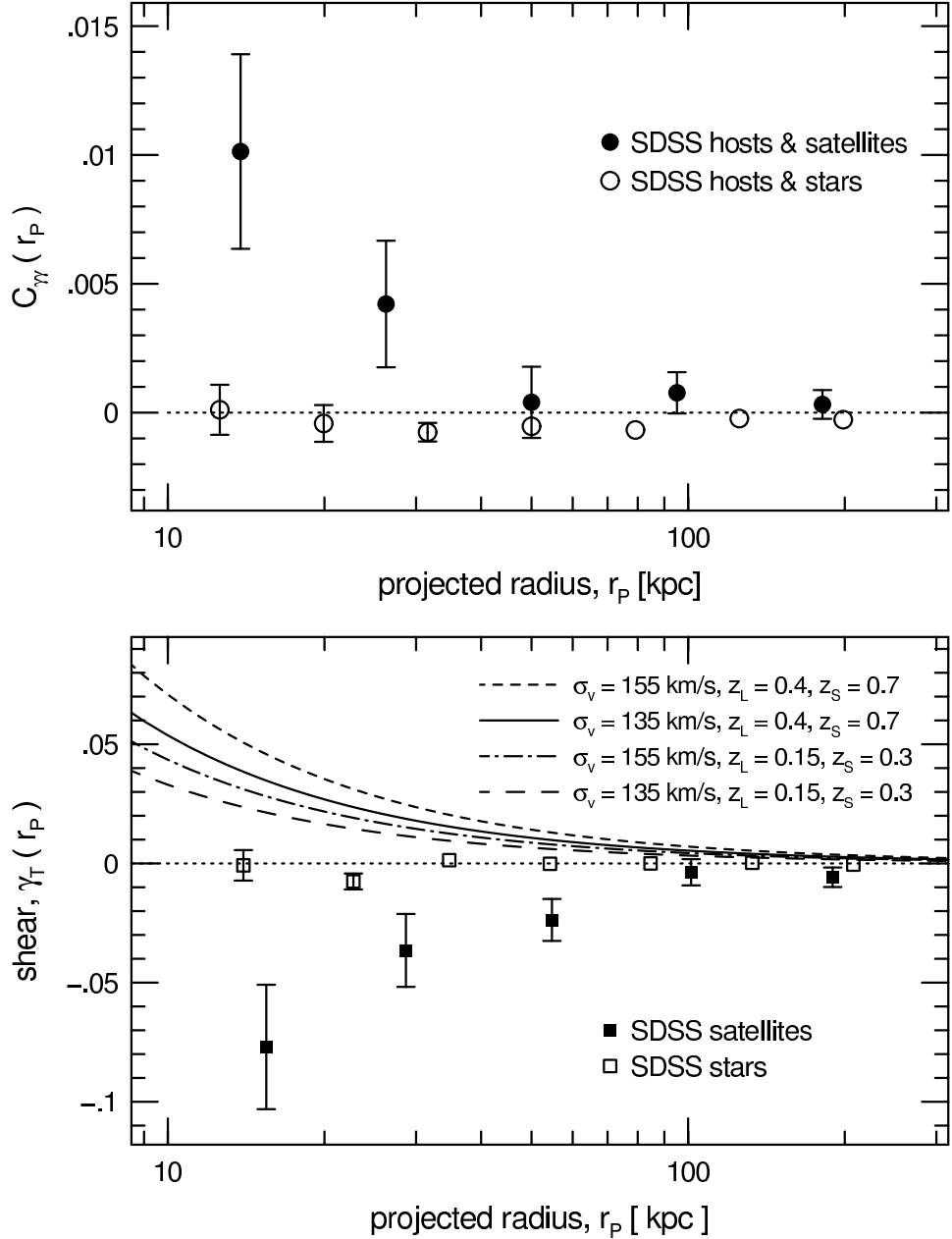}%
}
}
\vskip 0.1cm
\caption{Top panel: Two--point correlation function of host and satellite
galaxy image shapes (solid circles), as well as host and
star image shapes (open circles).  Positive values of $C_{\gamma \gamma}(r_p)$
indicate alignment of host and satellite images.  Bottom panel:
Mean tangential shear for the SDSS satellite
galaxies (solid squares) and stars (open squares). Negative
tangential shear indicates a systematic distortion in the radial direction.
Dotted line indicates $\gamma_T(r_P) = 0$.  Dashed, solid, and 
dot--dash lines show the theoretical tangential shear for isothermal sphere
lenses with velocity dispersion $\sigma_v$.  
}
\label{fig3}
\end{figure}

Since our satellites
are not randomly oriented with respect to their hosts, and
because a non--random orientation has potentially important implications
for GG lensing, we use the satellites to compute the
standard weak lensing quantity known as the mean
tangential shear, $\gamma_T$.
The tangential ellipticity for each satellite, $j$, is
computed as
$\gamma_j = \epsilon_j \cos(2 \alpha_j)$ where again
$\epsilon_j \equiv (a_j - b_j)/(a_j + b_j)$ and $\alpha_j$ is the
angle between the major axis of the satellite and the tangent to the
direction vector that connects the centroids of the host and satellite.
A simple,
unweighted mean of the individual values of $\gamma_j$ is used to
compute $\gamma_T$.  A positive value of $\gamma_T$ indicates tangential
orientation of the satellite images while a negative value of
$\gamma_T$ indicates radial orientation.
Solid squares in the bottom panel of Fig.~3 show
$\gamma_T$ for the SDSS satellites, computed
as a function of $r_P$.  Open squares show $\gamma_T$ for
stars with $16.5 \le r \le 18.5$ that are nearby to the hosts.
As expected from 
Fig.~2, the satellites have a negative value of $\gamma_T$
(most notably on scales $< 100$~kpc) and $\gamma_T$ for the stars
is consistent with zero.  
In particular, for
satellites with 
$10~{\rm kpc} \le r_P \le 50~{\rm kpc}$, the mean tangential shear is
$\gamma_T = -0.045 \pm 0.010$.
Although the sign of $\gamma_T$ is the opposite of what one expects
in gravitational lensing, the magnitude of $\gamma_T$ is 
comparable to what one would expect from galaxy--mass lenses with similar
impact parameters.  Shown in Fig.~3 for comparison are the
expected functions $\gamma_T(r_P)$ for four simple scenarios in which 
lens galaxies are modeled as
singular isothermal spheres with velocity dispersion
$\sigma_v$.  The lenses are located at redshift $z_L$ and the sources
are located at redshift $z_S$.  The values of $z_L$ and $z_S$ are similar to 
the actual median redshifts in current GG lensing studies, and
the values of $\sigma_v$ roughly span the range of values
that have been inferred for lens galaxies from GG lensing.

If a significant number of unidentified
satellite galaxies are present amongst the ``source'' galaxies in a
GG lensing data set, a substantial reduction of the 
true shear could result.  Contamination
by satellites is most likely to occur when
lenses and sources have been identified solely on the basis of apparent
magnitude (i.e., ``bright'' galaxies are identified as lenses and
``faint'' galaxies are identified as sources).
In these cases, the typical
difference in apparent magnitude between the ``sources'' and the ``lenses''
is of order 2 to 2.5 magnitudes, which is similar to the magnitude
difference between the hosts and satellites in our sample (see also
BN).  In addition, the vast majority of the SDSS
satellites
have apparent magnitudes that are fainter than the vast majority of the
SDSS hosts, so it is reasonable to estimate the degree to which bright
satellites might contaminate the GG lensing signal in a data set
that has a similar range of magnitude differences amongst its ``bright''
and ``faint'' galaxies.  In the magnitude range $16.5 \le r \le 18.5$, then,
we find that genuine satellite galaxies account for $\sim 10$\% to
 $\sim 15$\% of the {\it total} number of SDSS galaxies that surround the
host galaxies over scales $10~{\rm kpc} \le r_P \le r_{\rm max}$, where
$25~{\rm kpc} \le r_{\rm max} \le 250~{\rm kpc}$.  If we use a conservative
estimate of 10\% for the satellite contamination of the GG lensing signal,
then, over scales $r_P \lesssim 250$~kpc  our
observed elongation of 
satellites in the direction of
 their hosts reduces the true tangential shear for the
model lenses in Fig.~3 by an amount between 25\%$\pm$5\% (high--redshift
lens with $\sigma_v = 155$~km~sec$^{-1}$) and 40\%$\pm$10\% (low--redshift
lens with $\sigma_v = 135$~km~sec$^{-1}$).

\section{Discussion}


The cause of the elongation of the satellites in the direction of their hosts
is likely to be twofold.  On small scales, distortions caused by the 
gravitational interaction of the satellites with their hosts may
occur, leading to tidal streams.  We have examined the images
of $\sim 300$ of the brightest hosts and a handful ($< 10$) do
seem to show the presence of faint tidal streams that connect
the hosts and satellites.  On large scales, the orientation of
the satellites most
likely reflects the tendency for galaxies to form in rough alignment
along filaments.

The only previous study to which our work is directly comparable is
that of BN, who selected hosts and satellites in a manner similar to ours.
BN, however,
concluded that the tangential ellipticity of the
2dFGRS satellites was consistent with zero.  There are a number
of factors that contribute to this discrepancy.
First, our work is based on a larger number of
satellites (4289 vs.\ 1819).  The difference in the number of satellites
is caused largely by the fact that BN's satellites are much fainter than
the majority of the satellites used here.
Second, BN computed the tangential ellipticity over
a large aperture of radius 500~kpc, and we find that the preferential
alignment of the satellites is restricted to $r_P < 100$~kpc.
Third, the fraction of ``interlopers'' (i.e., galaxies
identified as satellites but which are not dynamically associated with
the host) is likely to be larger in BN than here.  This
is due to a combination of the facts that the radial velocity errors are
larger in the 2dFGRS than they are in the SDSS ($\sim 85$~km~sec$^{-1}$
vs.\ $\sim 25$~km~sec$^{-1}$) and that the interloper fraction increases
substantially with projected radius (e.g., Prada et al.\ 2003; Brainerd 2004b). 
Finally, the resolution of the satellite images used by BN is likely to
have been lower than the resolution of the SDSS satellite images used
here.  This is in part due to the fact that BN restricted their satellites
to be at least 7.6 times fainter than their host (so, on average, their 
satellites would have subtended a smaller angle on the sky than our
SDSS satellites).  In addition, the imaging of BN's satellites was based on
scans of glass plates with a pixel scale of $\sim 0.7''$, compared to 
a CCD pixel scale of $\sim 0.4''$ in the SDSS.

To compare to BN, we have converted SDSS magnitudes to the
$b_J$--band using the photometric transformation of Norberg et al.\ (2002),
$b_J = g + 0.155 + 0.152(g-r)$, and have identified isolated hosts
and their satellites in the DR4 using the same criteria as BN.  
Specifically, hosts
must be at least 7.6 times brighter than any other galaxy located within
$r_P \le 500$~kpc and $|dv| \le 1000$~km~sec$^{-1}$, and satellites consist
of all galaxies found within $r_P \le 500$~kpc and $|dv| \le 500$~km~sec$^{-1}$
of a host.  Implementing these criteria yields 1074~hosts
and 1467 satellites, from which we compute $P(\theta)$ and 
$P(\theta \le \theta_{\rm max})$.  When all satellites with
$r_P \le 500$~kpc are used, both $P(\theta)$ and 
$P(\theta \le \theta_{\rm max})$ are consistent with random distributions
($\chi^2$ rejection confidence level of 61.1\% and KS rejection confidence
level of 49.3\%).  When only those satellites with
$r_P \le 250$~kpc (i.e., the maximum $r_P$ for our
satellites) are used, the total number of satellites is reduced by $\sim 50$\%
and both $P(\theta)$ and $P(\theta \le \theta_{\rm max})$ remain
consistent with random distributions
($\chi^2$ rejection confidence level of 48.2\% and KS rejection confidence
level of 83.8\%).   Thus, when selected in a comparable
manner,
the SDSS galaxies yield
results that are consistent with BN's.  
In addition, 
we have applied our selection criteria from \S2 to the final data release
of the 2dFGRS, resulting in a sample of 2054 hosts and 2663 satellites.  A
preliminary analysis of the images of the satellites shows a
weak tendency for them to be elongated in the direction
of their
hosts;  a uniform distribution
for the 2dFRGS satellites is rejected 
at the 95.4\% confidence level by the $\chi^2$ test and at 
the 91.8\% confidence level by the KS test.  Averaged over scales 
$10~{\rm kpc} \le r_P \le 50~{\rm kpc}$, 
the mean tangential shear of the
2dFGRS satellites is $\gamma_T = -0.019 \pm 0.010$, in modest agreement
with our measurement for the SDSS DR4 satellites.  We will further 
explore the distortion of satellite images in both the SDSS and 2dFGRS,
including possible dependencies on host luminosity and spectral type,
in a future paper (Brainerd et al.\ 2006).

In conclusion, we find that
on scales $< 100$~kpc satellite galaxies are elongated in the 
direction of their
hosts. In addition, on scales $< 50$~kpc the images of the hosts
and their satellites are aligned with each other.
A decade ago the first
tentative detections of GG lensing yielded only noisy constraints
on the shear, and the presence of non--randomly oriented satellites amongst
the lensed sources
could be largely ignored.  Our results here, however, suggest that in the future
great care must be taken
to reject satellite galaxies (via, e.g., accurate photometric
redshifts) in order for ``precision shear''
observations of GG lensing
to result in truly precision constraints on the nature of
dark galaxy halos. 

\section*{Acknowledgments}

We are grateful to an anonymous referee for helpful suggestions
that improved the manuscript, and for support
under NSF contract AST-0406844.
Funding for the SDSS has been provided by the
A.\ P.\ Sloan Foundation, the Participating Institutions, NASA,
the NSF,
the US Dept.\ of Energy, the Japanese Monbukagakusho, and the Max
Planck Society. 
The SDSS is managed by the Astrophysical Research Consortium for
the Participating Institutions
(Univ.\ of Chicago, Fermilab, the Institute for Advanced Study, the
Japan Participation Group, Johns Hopkins Univ., Los Alamos National
Laboratory, the Max-Planck-Institute for Astronomy, the
Max-Planck-Institute for Astrophysics, New Mexico State Univ.,
Univ.\ of Pittsburgh, Princeton Univ., the US Naval
Observatory, and Univ.\ of Washington).

\end{document}